\documentclass[draft]{agujournal2019}
\usepackage{url} 
\usepackage[finalnew]{trackchanges} 
\usepackage{soul}

\usepackage{amsmath}
\usepackage{savesym}
\savesymbol{tablenum}
\usepackage{siunitx}
\draftfalse

\journalname{Space Weather}

\begin{document}

\title{Using gradient boosting regression to improve ambient solar wind model predictions}

%
%

\authors{R. L. Bailey\affil{1,2}, M. A. Reiss\affil{2,4}, C.~N.~Arge\affil{3}, C. M\"ostl\affil{2,4}, C.~J.~Henney\affil{5}, M.~J.~Owens\affil{6}, U.~V.~Amerstorfer\affil{2}, T.~Amerstorfer\affil{2}, A. J. Weiss\affil{2,7}, J.~Hinterreiter\affil{2,7}}

\affiliation{1}{Conrad Observatory, Zentralanstalt f\"ur Meteorologie und Geodynamik, 1190 Vienna, Austria}
\affiliation{2}{Space Research Institute, Austrian Academy of Sciences, Graz, Austria}
\affiliation{3}{Heliophysics Science Division, NASA Goddard Space Flight Center, Greenbelt, MD 20771, USA}
\affiliation{4}{Institute of Geodesy, Graz University of Technology, Graz, Austria}
\affiliation{5}{Air Force Research Laboratory, Space Vehicles Directorate, Kirtland AFB, NM, USA}
\affiliation{6}{Space and Atmospheric Electricity Group, Department of Meteorology, University of Reading, Reading, UK}
\affiliation{7}{Institute of Physics, University of Graz, Graz, Austria}

\correspondingauthor{Rachel Bailey}{rachel.bailey@zamg.ac.at}




\begin{keypoints}
\item We present a machine learning approach to use in place of the WSA model for predicting solar wind speed near Earth.
\item The tool is validated on one whole solar cycle and outperforms established models and a baseline recurrence model.
\item This study presents a fast and well-validated open-source contribution to the field of ambient solar wind modeling and prediction.
\end{keypoints}

\begin{abstract}
Studying the ambient solar wind, a continuous pressure-driven plasma flow emanating from our Sun, is an important component of space weather research. The ambient solar wind flows in interplanetary space determine how solar storms evolve through the heliosphere before reaching Earth, and especially during solar minimum are themselves a driver of activity in the Earth's magnetic field. Accurately forecasting the ambient solar wind flow is therefore imperative to space weather awareness. Here we present a machine learning approach in which solutions from magnetic models of the solar corona are used to output the solar wind conditions near the Earth. The results are compared to observations and existing models in a comprehensive validation analysis, and the new model outperforms existing models in almost all measures. In addition, this approach offers a new perspective to discuss the role of different input data to ambient solar wind modeling, and what this tells us about the underlying physical processes. The final model discussed here represents an extremely fast, well-validated and open-source approach to the forecasting of ambient solar wind at Earth. 
\end{abstract}

\section{Introduction} \label{sec:intro}

It was only in the 1970s, through the view of X-ray telescopes on the NASA Skylab satellite, that the dark patches in the Sun's polar regions were identified as coronal holes. Over the past decades, it has become clear that these are areas of open magnetic field lines, along which the solar plasma leaving the Sun in a continuous flow is accelerated to supersonic speeds into the vast reaches of our solar system. This fast component of the ambient solar wind flow and the magnetic field embedded within it corotate with the Sun in an ever expanding spiral, the Parker spiral. Understanding of the ambient solar wind is important in space weather forecasting to determine the evolution and possible impact of solar storms, particularly as they influence the evolution of transient events such as coronal mass ejections (CMEs), the catalysts for the strongest geomagnetic storms, on their path from Sun to Earth~\cite{temmer11, owens13a}. Ambient solar wind flows are themselves also an important driver of recurrent geomagnetic activity at Earth~\cite{Zhang2007, Verbanac2011, Nakagawa2019}, determining critical properties in interplanetary space such as the solar wind speed, magnetic field strength and orientation~\cite{luhmann02}. A clear picture and understanding and accurate modeling of the ambient solar wind are therefore essential in all aspects of space weather research.

In today's ambient solar wind modeling approaches, the solar surface, corona and inner heliosphere are treated as a connected system to simulate the dynamics of the ambient solar wind flow from the Sun to the vicinity of Earth. These coupled modeling approaches commonly span the range from 1 solar radius ($R_\odot$) to Earth at 1 AU, with the coronal model from 1 $R_\odot$ to 5 $R_\odot$ (or $30$ $R_\odot$) and the heliospheric model spanning the range from 5-30 $R_\odot$ to 1 AU. Despite the discovery of an empirical relationship between the configuration of open magnetic field lines on the Sun and the solar wind properties measured at the Sun-Earth Lagrange Point 1 (L1) near the Earth~\cite{wang90}, it still proves challenging to develop and optimize empirical techniques for specifying the solar wind conditions at the inner boundary of the heliospheric domain~\cite{arge00,riley01,arge04}. Well-known empirical relationships in this context are the Wang-Sheeley (WS) model~\cite{wang90}, Distance from the Coronal Hole Boundary (DCHB) model~\cite{riley01} and the Wang-Sheeley-Arge (WSA) model~\cite{arge03}. More sophisticated three-dimensional magnetohydrodynamic (MHD) codes such as CORHEL~\cite{Linker2016}, LFM-Helio~\cite{Merkin2016}, SIP-CESE~\cite{Feng2015} and COIN-TVD MHD~\cite{Shen2018} are also used, with further examples being the Magnetohydrodynamics Algorithm outside a Sphere~\cite{linker99}, Enlil~\cite{odstrcil03}, the Space Weather Modeling Framework~\cite{toth05}, and the recently developed European Heliospheric Forecasting Information Asset~\cite{pomoell18}. Besides these MHD models, other modeling approaches based on empirical relationships and statistics~\cite<e.g.>{owens20} have also been developed. In the world of machine learning, early work of \citeA{Wintoft1997} and \citeA{Wintoft1999} developed neural networks to forecast the solar wind speed near Earth using values from coronal magnetic models, and similar ideas were developed further in \citeA{Liu2011}, \citeA{Yang2018} and \citeA{Chandorkar2020}.

From the available models, the WSA model approach enjoys widespread use in operational settings in the Met Office and NOAA's Space Weather Prediction Center, primarily in the prediction of high-speed solar wind streams \cite{macneice09b,owens05,reiss16,reiss19}. It also has applications for many different scientific problems such as the evolution of CMEs in interplanetary space and their subsequent arrival at Earth \cite{mays15,riley18,scolini19,taktakishvili09,verbeke19,wold18}. Prediction of solar energetic particle events \cite{luhmann17,macneice11} and studies evaluating the interaction between the solar wind and planetary magnetospheres \cite{dewey15} have also been carried out using the WSA model, highlighting its importance in current research.

Despite all the recent developments, ambient solar wind forecasting models still only outperform 27-day solar wind persistence by a small margin, often with uncertainties in the timing of high-speed streams of roughly one day~\cite{Allen2020,devos14, hinterreiter2019,jian11,kohutova16,owens08,owens13b,reiss16}. One large source of uncertainty is in the modeling of the topology of the coronal magnetic field, which the WSA model takes as input, as we typically only have data for one side of the Sun. This is tackled in one way using the Air Force Data Assimilative Photospheric Flux Transport or ADAPT model \cite{Arge2010,Henn2012,hickmann15}, which provides multiple realizations of the possible magnetic field on the far side of the Sun. As can be seen by the many and varied approaches used, forecasting ambient solar wind conditions remains a challenge. 

Continued efforts are needed to improve our capabilities for predicting conditions in the ambient solar wind. Here we improve ambient solar wind forecasts by coupling machine learning techniques to modeling results from the outer boundary condition of coronal magnetic models. Machine learning techniques are undergoing a renaissance and are growing in popularity among both the public and the space weather community as a new tool for current challenges, with promising results. An overview of recent developments in machine learning methods applied to the topics of space weather and the heliosphere can be found for example in~\citeA{Camporeale2019}, with studies tackling problems such as forecasting geomagnetic indices and automated solar image classification. So far, however, beside the application of statistical techniques \cite{BussyVirat2014, Riley2017}, there have been few studies on ambient solar wind forecasting using machine learning. Here we will couple machine learning with established solar wind methods in combination with the ADAPT output. ADAPT produces an ensemble of solutions, but strategies looking at how to best combine these ensembles are still missing. Machine learning offers an interesting opportunity to find ways to optimally combine magnetic map solutions with the aim to transition the research to operational usage.

\begin{figure*}
\begin{center}
\includegraphics[width=\textwidth]{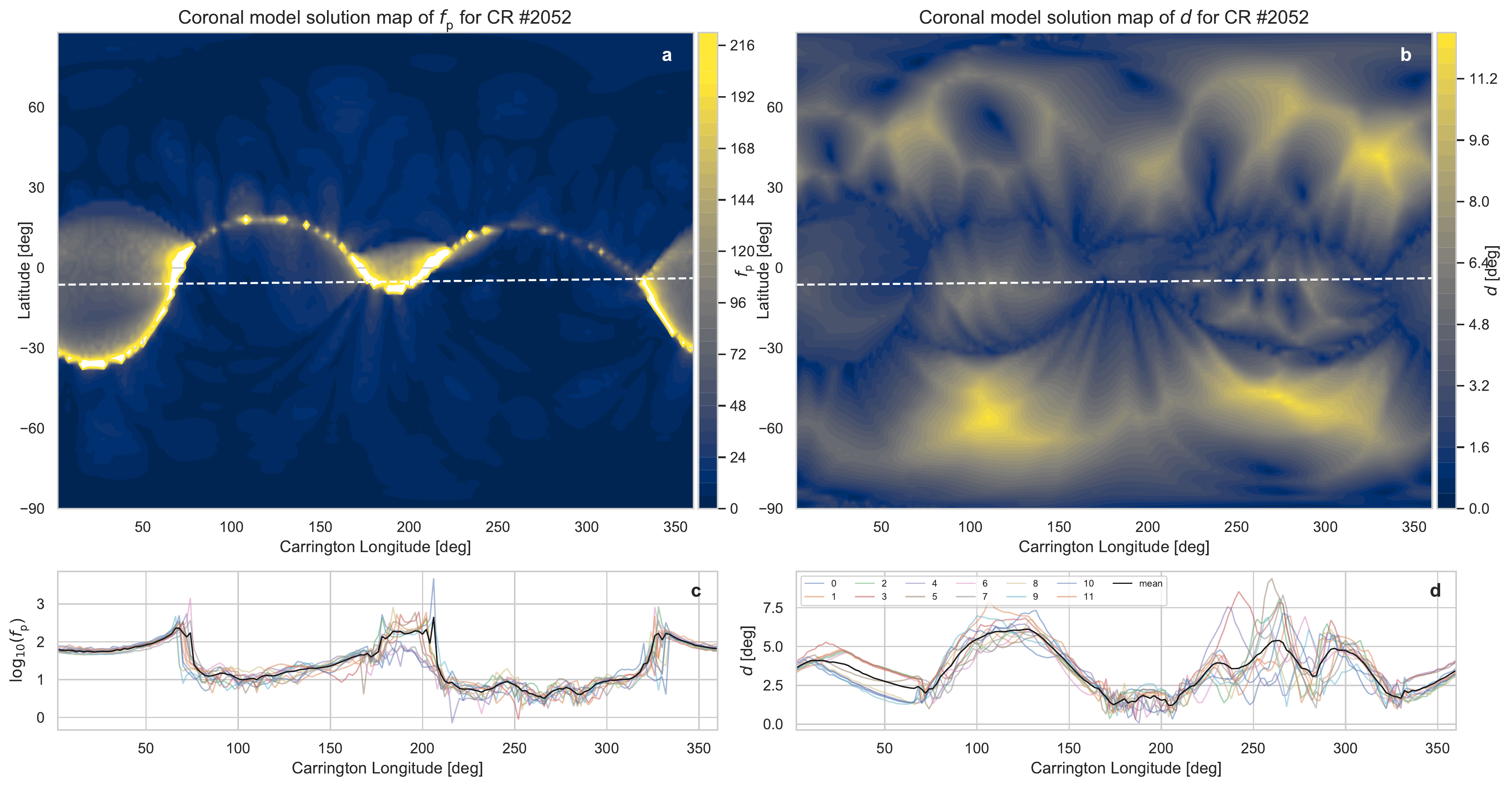}
\caption{The flux tube expansion factor $f_\text{p}$ (left) and distance to the coronal hole boundary $d$ (right) are shown for one Carrington rotation (CR \#2052). (a-b) Coronal model maps for each variable at the outer boundary of the coronal model ($5 \ R_\odot$) with the sub-Earth track drawn in with a dashed white line. (c-d) The values extracted along the sub-Earth track. Output from all 12 ADAPT realizations are shown in colour, while the black line depicts the mean. The log of $f_\text{p}$ is shown so that the behaviour is more easily interpretable in the plot, but the value did not need to be scaled for the model training. Note that, from a time perspective, time is running in the opposite direction with decreasing Carrington Longitude as the Sun rotates from left to right according to the map.}
\label{fig:fp_d}
\end{center}
\end{figure*}

The machine learning method we apply in this study is a Gradient Boosting Regressor (GBR), which is a powerful technique that builds a single estimator from a collection of weak learners called decision trees~\cite{Friedman2001}. We use the GBR to predict the ambient solar wind bulk speed at Earth using the magnetic topology solutions at the outer boundary of the coronal magnetic model. Figure \ref{fig:fp_d} shows a depiction of the WSA coronal model solutions for one Carrington rotation and the variables extracted for model training. The top part represents maps of the flux tube expansion factor $f_\text{p}$ and distance to the coronal hole boundary $d$ computed at a reference height of $5 R_\odot$. The lower part shows each variable extracted from the map along the sub-Earth track, which is the path along which a projection of the Earth passes during the Carrington rotation. All 12 ADAPT realizations can be included in the machine learning approach to account for uncertainties in the magnetic modeling. Although the focus is on the improvement of the modeling capabilities, we will show that, through the training of our model, we can study the fundamental properties of solar wind models, which provides clues to the underlying physical mechanisms responsible for accelerating the solar wind.

We present the application of machine learning techniques to complement and inform established solar wind modeling approaches. The machine learning model is trained on 14 years of past data (1992 till 2006) and tested on one full solar cycle length of 11 years (2006 till 2017) from the recent Solar Cycle 24. We present a comprehensive validation analysis of the machine learning model (GBR) solutions and compare the results to commonly applied ambient solar wind models and reference baseline models, including a 27-day persistence model, which uses the approximation that the conditions in the ambient solar wind flow repeat after each Carrington rotation. This is followed by a discussion of the importance and then outline possible future investigations. This is the first study in which all ADAPT realizations are coupled with a machine learning approach.

\section{Methods}

\subsection{Magnetic models of the Sun}

In this section, we present the numerical framework for reconstructing the global magnetic field topology in the solar corona. Our framework relies on magnetic maps of the photospheric magnetic field from the Global Oscillation Network Group from the National Solar Observatory. Specifically, we make use of the ADAPT (Air Force Data Assimilative Photospheric Flux Transport) \cite[Air Force Data Assimilative Photospheric Flux Transport]{Arge2010,Henn2012,hickmann15} model. ADAPT is a flux transport model (e.g., includes differential rotation and meridional flows) to provide an ensemble of estimates of the global spatial distribution of the solar photospheric magnetic field. The ADAPT model produced ensembles of twelve realizations representing the uncertainty driven by supergranulation \cite{Worden2000}, which means that each ADAPT realization depicts a slightly different magnetic field configuration depending on the assumptions that are made for the magnetic field on the far side of the Sun.

Based on the twelve different ADAPT realizations, we reconstruct the global coronal magnetic field topology using the commonly-employed Potential Field Source Surface (PFSS)~\cite{altschuler69,schatten69} and Schatten Current Sheet (SCS)~\cite{schatten71} model combination. The well-established PFSS model computes the potential magnetic field solution in the solar corona with an outer boundary condition that the magnetic field is purely radial at the source surface at $2.5 \,R_\odot$. The SCS model in the modeling domain between $2.5$ and $5$ $R_\odot$ accounts for the latitudinal invariance of the radial field as observed by the \textit{Ulysses} mission~\cite{wang95}. 

We compute magnetic properties such as the areal expansion factor and the distance to the coronal hole boundary from the modeled global magnetic field configuration. The areal expansion factor $f_\text{p}$ is the rate at which the flux tube expands between the photosphere and a reference height in the corona~\cite{wang90}:
\begin{equation}
f_{\text{p}} = \left( \frac{R_\odot}{2.5 \ R_\odot} \right)^2 \frac{\left|\textbf{B}(R_\odot,\theta_0,\phi_0)\right|}{\left|\textbf{B}(2.5 \ R_\odot,\theta_1,\phi_1)\right|}.
\end{equation}
$\theta$ is the longitude and $\phi$ the latitude, where the indices 0 and 1 represent the longitude/latitude at the solar surface and 2.5 solar radii, respectively. We trace the field lines down from 5 $R_\odot$, then calculate the flux tube expansion factor between 2.5 $R_\odot$ and the solar surface in line with the definition given in~\citeA{wang90}.

The distance to the coronal hole boundary $d$ refers to the great circle distance that an open-field footpoint in the photosphere lies from the nearest coronal hole boundary. It is calculated at the footpoint of the field line on the solar surface, and this value is then mapped along the magnetic field to a distance of 5 $R_\odot$. The underlying idea of $d$ is that the solar wind is slow near coronal hole boundaries and fast inside regions of open field topology~\cite{riley01}.

From the ensemble solutions of ADAPT, we obtain a set of twelve different results for all the magnetic properties computed. Besides the expansion factor and the distance to the nearest coronal hole boundary at the sub-Earth point, we also study the usefulness of the magnetic field strength at the photospheric footpoint, $B_{\text{phot}}$, and the outer boundary of the coronal model, $B_{\text{cor}}$.

As a reference baseline model, we use the traditional WSA approach~\cite{arge03} for specifying the conditions in the solar wind near the Sun. The WSA model used here is a combination of the expansion factor and the distance from the nearest coronal hole boundary. Specifically, the WSA relation used in this study is given by 
\begin{equation}
v_{\text{wsa}} (f_\text{p}, d) = c_1 + \frac{c_2}{\left(1 + f_\text{p}\right)^{c_3}} \left\{ c_4 - c_5 \ \exp \left[ {-\left(\frac{d}{c_6}\right)}^{c_7}\right] \right\}^{c_8},
\label{eq:wsa}
\end{equation} 
where $c_i$ are the model coefficients, for which we use the following settings: $c_1=250~\si{km.s^{-1}}$, $c_2=650~\si{km.s^{-1}}$, $c_3=0.19$, $c_4=1$, $c_5=0.8$, $c_6=3^{\circ}$, $c_7=1.75$ and $c_8 =3$~\cite{arge03}. 

To map the solar wind solutions near the Sun to the Earth, we employ the Heliospheric Upwind eXtrapolation model \cite<HUX,>{riley11b}, which simplifies the fluid momentum equation as much as possible. Furthermore, the HUX model solutions match the dynamical evolution explored by global heliospheric MHD codes reasonably well while having low computing requirements~\cite{owens20}. In this way, we can efficiently study the results and implications of the ensemble ADAPT realizations. For more details on the HUX model, we would like to refer the reader to~\citeA{riley11b} and~\citeA{owens17}. 

\subsection{Application of Machine Learning}

We use a machine learning approach to predict the solar wind speed at L1 from the output of the coronal magnetic model. The steps worked through in order to train and finalise the machine learning model can be summarised as follows. To begin with, the data at a height of $5 R_\odot$ is extracted from the coronal magnetic model along the sub-Earth track to produce a time series. Next, this data set is split into training and testing data, with 14.5 years being used to train the model, and one full solar cycle length (11 years) is left to test the model on. The model that will be trained is a Gradient Boosting Regressor (GBR), and an initial iterative training through sets of machine learning-specific parameters is carried out on the data set to determine those best-suited for this specific problem. With the model parameters determined, the model is first trained on the full data set before feature selection is carried out to reduce the number of input features. This produces a set of trained models, which we compare using validation metrics. The optimal model is saved for later use.

The results extracted from the coronal magnetic model are the properties $f_\text{p}$, $d$, $B_{cor}$ and $B_{phot}$ calculated along the sub-Earth track, i.e. the path the Earth traces through the rotation projected along the solar surface. There were 12 coronal magnetic model solutions per day based on 12 different ADAPT solutions. The sub-Earth track was extracted from each of these, and all ADAPT solutions were included. The data was extracted in a way equivalent to an operational setting, in which the time series is updated once per day with the newest coronal magnetic model results (this is equivalent to 6-7 time steps per day). In order to compare the model solutions with observations, the solar wind solutions, which were not periodic due to differing Carrington rotation lengths, were interpolated to periodic time stamps corresponding to a longitudinal resolution of 2 degrees (3.64 hours). Figure~\ref{fig:fp_d} shows the full coronal model maps and one path extracted along the sub-Earth track, while Fig.~\ref{fig:fp_d_time} shows both variables extracted to form a time series.

In machine learning language, the \textit{features} are the magnetic model properties described above, and the \textit{target} is the solar wind speed $v_\text{sw}$ in the near-Earth environment. The model is trained to be able to produce the target from any given set of features.
The solar wind speed is taken from the OMNI hourly data set. In models developed to forecast the ambient solar wind speed from magnetic model results, there is usually a heliospheric model to account for expansion times from the outer boundary of the coronal model to L1. 

In this approach, the use of a singular defined time shift is avoided and instead the variables over the past 2-6 days are taken as input, meaning that at a resolution of 3.64 hours, there are 29 time-shifted values of four properties taken from 12 possible ADAPT realizations, resulting in a total of 1,392 magnetic model features. In addition to the magnetic properties, a solar wind bulk speed persistence variable, $v_\text{pers}$, was included, under the assumption that the variation in the solar wind speed repeats itself every 27.27 days. Three persistence features were included: the solar wind speed 26, 27 and 28 days before the target forecast, totaling 1,395 features. Contrary to many other machine learning approaches, the features do not need to be scaled for a GBR. As in decision trees, the model picks points in the range of features at which to split nodes to form branches, leading to different output values at the leaves. The locations of these splits are insensitive to feature scaling.

In order to test the accuracy of the final model, the input data is split into training and testing data for the sake of fair testing and model validation. The test data set constitutes an entire average solar cycle length, none of which has been seen by the model in the training data set. Keeping an entire set of varied solar cycle conditions unseen and using this for validation ensures we evaluate the model's ability to extrapolate to new behaviour. The training data set covers 14.5 years, most of solar cycle 23, and runs from May 1992 till October 2006, while the test set covers the 11 years from October 2006 till October 2017, which is the end of solar cycle 23 and a large part of 24.

The machine learning method chosen for this study relies on Gradient Boosting Regressors (GBRs), and the specific implementation applied here is the Python version of XGBoost~\cite{Chen2016}. The GBR was chosen from an initial study with a group of regressors including others such as Random Forest Regressors. GBRs are based on an ensemble of decision trees, which on their own are weak learners, but in a forest form a powerful prediction tool. \remove{In a field quickly becoming saturated with neural nets, we chose this method to offer a different approach with a more robust method of training and easier interpretation.} The term gradient boosting refers to the gradient descent technique implemented for optimised fitting. For more details on the algorithm, see~\citeA{Friedman2001}. A summary of the general usage of GBRs can be found in~\citeA{Natekin2013}. 

\change{There are many benefits to GBRs, which include the ease of use and interpretation coupled with fast computation times.}{Gradient boosting regressors and neural networks have their distinct advantages and disadvantages. In comparison to neural networks, GBRs have the benefit of being easier to interpret because the mechanisms behind their predictions (i.e. decision trees) are well-understood. They also have easier and more intuitive implementation and tuning, leading to faster model development than with neural networks. On the other hand, the trade-off for easier implementation is that they require a lot of data preparation beforehand. One of the strengths of deep learning neural networks is that the data can often be provided with minimal handling because the network can learn many relationships without them being explicitly given. This is not as easy in gradient boosting regression, so in the following we clarify all the steps applied to the data to prepare it for model training. We choose a GBR over a more complex neural network (e.g. recurrent or convolutional neural networks) for the simpler implementation and interpretation coupled with powerful learning capabilities, and to tackle the problem of ambient solar wind forecasting in a new way. We compare the results to the output from neural networks applied in} \citeA{Yang2018} and \citeA{Chandorkar2020} \add{in the discussion.}

Hyperparameter tuning is carried out for the GBR for a specific set of features, in which a grid search is used to evaluate the best combination of GBR-specific input parameters\remove{ (number of estimators, learning rate, maximum depth of nodes and L1 regularisation terms)}. This is done by exploring every point in the 4-D parameter space within certain parameter ranges and training a model on each. The combination with the best predictive ability according to the selected scoring measure is used for final training. \add{Using the results of the grid search, we built a model of 600 decision trees - any more and it would tend towards overfitting. The maximum depth of each tree was set to 3, the learning rate to 0.3 and the L1 regularisation term to 20.} \change{Different loss functions were also tested with the model, with the model default RMSE proving most useful.}{A collection of loss functions commonly applied to regression problems were also compared: the root-mean-square error (RMSE), root mean square log error (RMSLE), mean absolute error (MAE) and the mean pseudo-Huber error (MPHE). The metrics MAE and MPHE, in particular, are useful for training datasets with many outliers that can bias training. Outliers were not a problem in our dataset, and the RMSE proved to be the most robust metric. In practice, however, models trained with either RMSE, MAE or MPHE as the loss metric had near-equivalent accuracy.}

The training data set is shuffled and provided to the model with an 80/20 split between train and validation data. Through stratified 5-fold cross-validation, the validation data set is kept aside and used during training to choose the best option among five models\add{, and early stopping is used to return the best model before overfitting occurs}.

\subsection{Feature selection} 
In machine learning, feature selection describes the act of reducing or adjusting the input features to optimise the final model, as well as reduce time needed for training. Training the model on the full set of 1,395 features with around 60,000 data points each is computationally heavy, and so the first step in feature selection is in finding a reduced set of variables that produces a model with an equal level of accuracy. A straightforward approach, in which the input variables over time are reduced so that only values at every -2, -3, -4, -5 and -6 days are taken, performs just as well or better than the feature list with all possible time steps, and the number of features drops to 288.
More detailed feature selection is carried out through different methods, firstly using an evaluation of the feature importances, which the trained model provides directly in terms of percentages for each individual feature, and then through training of the model on individual variables and different combinations thereof.
More details on feature importances in GBRs can be found in~\citeA{Hastie2009}.

\begin{figure*}
\begin{center}
\includegraphics[width=\textwidth]{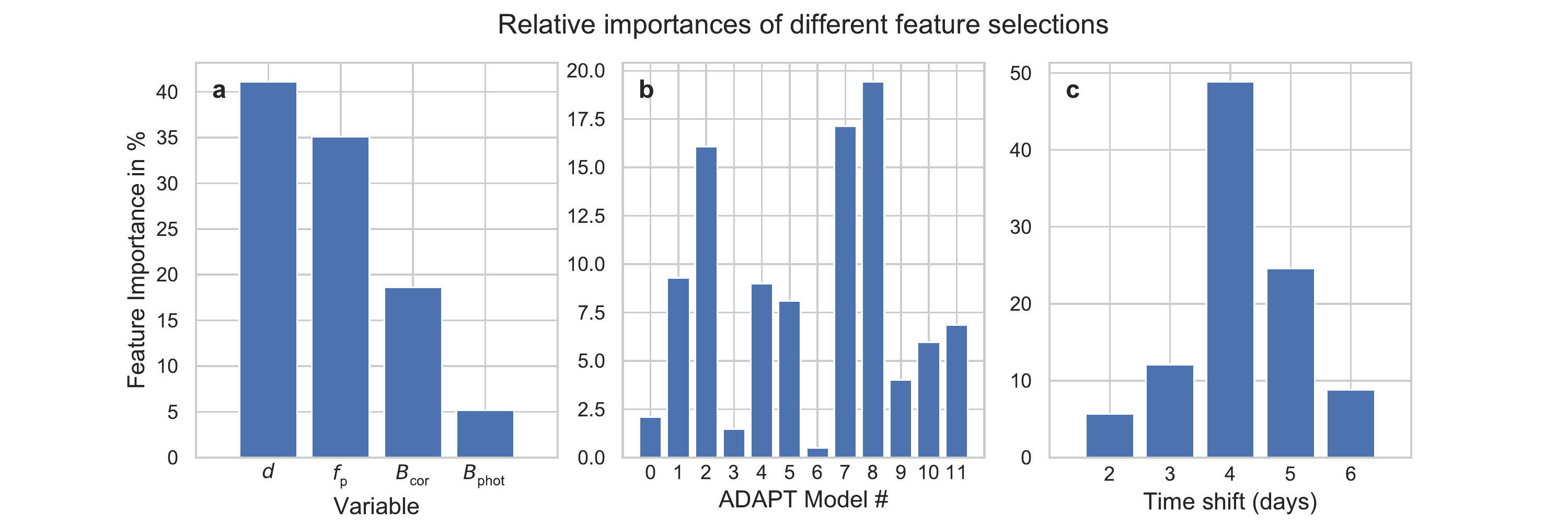}
\caption{Importance of features when training the machine learning model. The feature importances of all features according to either (a) the variable used, (b) the ADAPT model realization the variable was taken from, or (c) the time shift applied to the variable.}
\label{fig:feature_imps}
\end{center}
\end{figure*}

Figure~\ref{fig:feature_imps} shows an evaluation of all feature importances with the features divided into groups. This shows the importance of each of the four variables to the model, followed by the relative importance of each ADAPT model realization, and lastly the importance each time shift (grouped into days) has for the model training. This plot largely follows expected patterns: $d$ and $f_\text{p}$ are clearly the most important properties as expected from the WSA approach, and among the ADAPT ensemble members some are notably better at predicting than others in this combination. The other two variables related to the coronal and photospheric magnetic fields do not contain enough information to carry an ambient solar wind model on their own merits and can be excluded. Among the time-shifted variables, the most important are around a shift of -4 days, which is the average time it takes for solar wind leaving the corona to reach the Earth, although data at other time shifts also play into it. These feature importances were evaluated without the influence of the statistically similar persistence solar wind speed $v_\text{pers}$, which trained alongside the others makes up almost 20\% of the model variance.

The best model, deduced from the feature importances, is one that uses both $f_\text{p}$ and $d$ with $v_\text{pers}$, with input from three ADAPT models (realization \#s 2, 7 and 8) and the number of time-shifted features reduced to only one per day, a total of 33 features. This performs better than one trained on all features, which is more likely overtrained -  $B_{cor}$ and $B_{phot}$ in particular seemed to be very situation-specific, and the model performed better on new data without them.
Once the best selection of features has been determined, the trained model is saved as a Python pickle file for future use and applied to the test data set for the validation analysis.

\subsection{Validation analysis}

Here we assess the correspondence between the predicted and observed time series of solar wind bulk speed in three different ways. Firstly, we present the validation analysis in terms of a continuous variable validation based on simple point-to-point comparison metrics. Secondly, we study the skill in terms of binary metrics, where we label each time step in the predicted and observed timeline as an event/non-event based on a selected threshold value. Thirdly, we complement the discussed validation techniques with an event-based approach where periods of enhanced solar wind speed in observations and forecasts are automatically detected and compared against each other. This validation analysis is based on previous research as discussed in~\citeA{owens18} and~\citeA{reiss20}. 

\subsubsection{Point-to-point error functions}

The predictive abilities of ambient solar wind models are assessed most easily by comparing the observed solar wind time series to ambient solar wind model solutions. A straightforward way is to compare the underlying statistical distributions in terms of standard measures such as mean, median, and standard deviation. These basic statistical measures already contain essential information on the tendency of the model to over- or underestimate the observed solar wind conditions. We complement these basic measures with established error functions. We compute different error functions for continuous variables such as the mean error,
\begin{equation}
\text{ME} = \frac{1}{n} \sum_{k=1}^n (f_k - o_k) = \bar{f} - \bar{o},
\end{equation}
where $(f_k,o_k)$ is the $k$-th element of $n$ total forecast and observation pairs. Here the ME is simply the difference between the average prediction and the average observation. Furthermore, we compute the mean absolute error, 
\begin{equation}
\text{MAE} = \frac{1}{n} \sum_{k=1}^n |f_k - o_k|.
\end{equation}
The MAE is the arithmetic mean of the absolute differences between the prediction and the observation pairs. It represents the typical magnitude for the prediction error. Similar to the MAE, the root-mean-square error, 
\begin{equation}
\text{RMSE} = \sqrt{\frac{1}{n} \sum_{k=1}^n (f_k - o_k)^2},
\end{equation}
is the mean squared difference between prediction and observation value pairs. The RMSE is an estimate of the magnitude of the forecast error being more sensitive to outliers than the MAE. The last error measure is the skill score:
\begin{equation}
\text{SS} = 1 - \frac{\text{MSE}_\text{pred}}{\text{MSE}_\text{ref}},
\end{equation}
which compares the mean-squared errors (MSE) between the prediction and reference. These measures are equal to zero in the case that the forecast errors are equal to zero (that is $f_k =o_k$) and increase with increasing disagreement between predictions and observations. Although strictly not an error function per definition, we complement the error functions with the Pearson correlation coefficient. 

\subsubsection{Binary metrics}

Aside from approaches assessing the magnitude of the prediction error at every time step in comparison to the observed value, we use another approach where each time step in the solar wind time series is labelled in binary values as an event or non-event. A threshold value is selected, and all values above the threshold are defined as events while those below are non-events. The advantage of this method is that slow and fast solar wind, of which fast solar wind is of more interest to end-users, are not given equal importance~\cite{owens18}. It is also generally insensitive to outliers, which are irrelevant for a forecast but can greatly affect point-to-point error measures.

The definition of events and non-events in the solar wind time series uses a selected threshold value~\cite{reiss20} of $450$ km/s.
We cross-check events and non-events in the predicted and observed solar wind time series and count the number of hits (true positives; TPs), false alarms (false positives; FPs), misses (false negatives; FNs) and correct rejections (true negatives; TNs), which are listed in a contingency table. A hit is a correctly predicted event, while a miss is an event that is observed but not predicted. In contrast, a false alarm is a predicted event that was not observed and a correct rejection is a correctly predicted non-event. The total counts are summarised in the so-called contingency table, and from these we compute a set of skill measures such as the \emph{true positive rate} $\text{TPR} = \text{TP}/(\text{TP} + \text{FN})$, \emph{false positive rate} $\text{FPR} = \text{FP}/(\text{FP} + \text{TN})$, \emph{threat score} $\text{TS} = \text{TP}/(\text{TP} + \text{FP} + \text{FN})$, \emph{bias} $\text{BS} = (\text{TP} + \text{FP})/(\text{TP} + \text{FN})$, and \emph{true skill statistics} $\text{TSS} = \text{TPR} - \text{FPR}$. The threat score is a measure of the overall performance of a model defined between 0 and 1 (best performance), while the bias depicts how well the model tends towards overpredicting (BS $>$ 0) or underpredicting (BS $<$ 0) events. A meaningful skill measure is the true skill statistics (TSS), which is defined in the range $[-1,1]$. A perfect prediction model would have a value of 1 (or -1 for perfectly inverse predictions), and a TSS of 0 indicates no skill. The advantage of the TSS is that it combines all elements of the contingency table, and is unaffected by the proportion of predicted and observed events~\cite{hanssen65,bloomfield12,wilks11}.

The receiver operator characteristic (ROC) curve neatly summarises the predictive capabilities for a range of different event thresholds. The curves illustrate how the number of correctly predicted events (TPR) varies with the number of incorrectly predicted non-events (FPR).

\subsubsection{Event-based validation}

The interpretation of simple point-to-point comparisons as described above can be misleading due to a lack of knowledge on the extent of uncertainties related to timing errors~\cite{owens05,macneice09a,macneice09b}. In particular, this is the case when large variations in the solar wind time series are generally well predicted, but the arrival times differ between prediction and observations. The use of an event-based validation technique is commonly applied to account for the uncertainties in the arrival times. More specifically, validation analysis on the example of solar wind bulk speed is carried out in three steps. First, events of enhanced solar wind speed, also called high-speed enhancements (HSEs) in forecast and observation data are defined as periods exceeding a certain threshold. Next, HSEs detected in the solar wind measurements are paired with HSEs in the predictions and each event pair is labelled as a hit, false alarm, or miss. The predictive abilities of the model are determined using the validation summary variables. A more detailed description of the steps discussed above applied to solar wind speed measurements is given in~\citeA{reiss16} in Section 3.2. 

\section{Results}\label{sec:analysis}

\begin{figure*}
\begin{center}
\includegraphics[width=\textwidth]{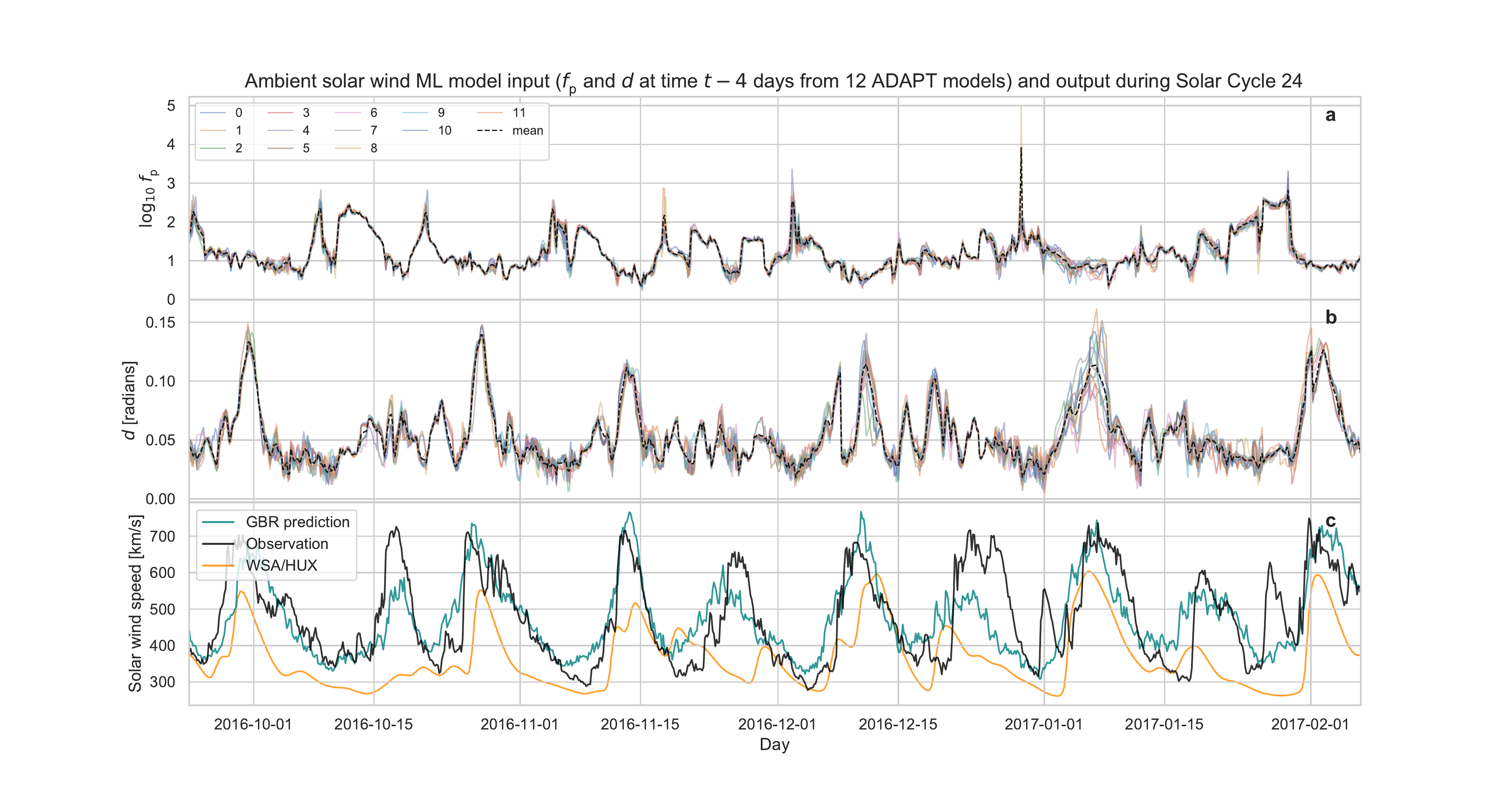}
\caption{Input (a-b) and output (c) from ambient solar wind model. Variation in flux tube expansion factor $f_\text{p}$ (a) and distance to the coronal hole boundary $d$ (b) over five Carrington Rotations (CR 2182-2187) at a time shift of $t-4$ days relative to the time step of the final solar wind speed prediction. Output from all 12 ADAPT realizations are shown in colour, while the black line depicts the mean. (c) Solar wind speed predicted from the machine learning model (teal) and the WSA/HUX model (orange) plotted against the observed solar wind speed (black). Predictions are during solar minimum in solar cycle 24, on which the model was tested.}
\label{fig:fp_d_time}
\end{center}
\end{figure*}

\begin{figure*}
\begin{center}
\includegraphics[width=\textwidth]{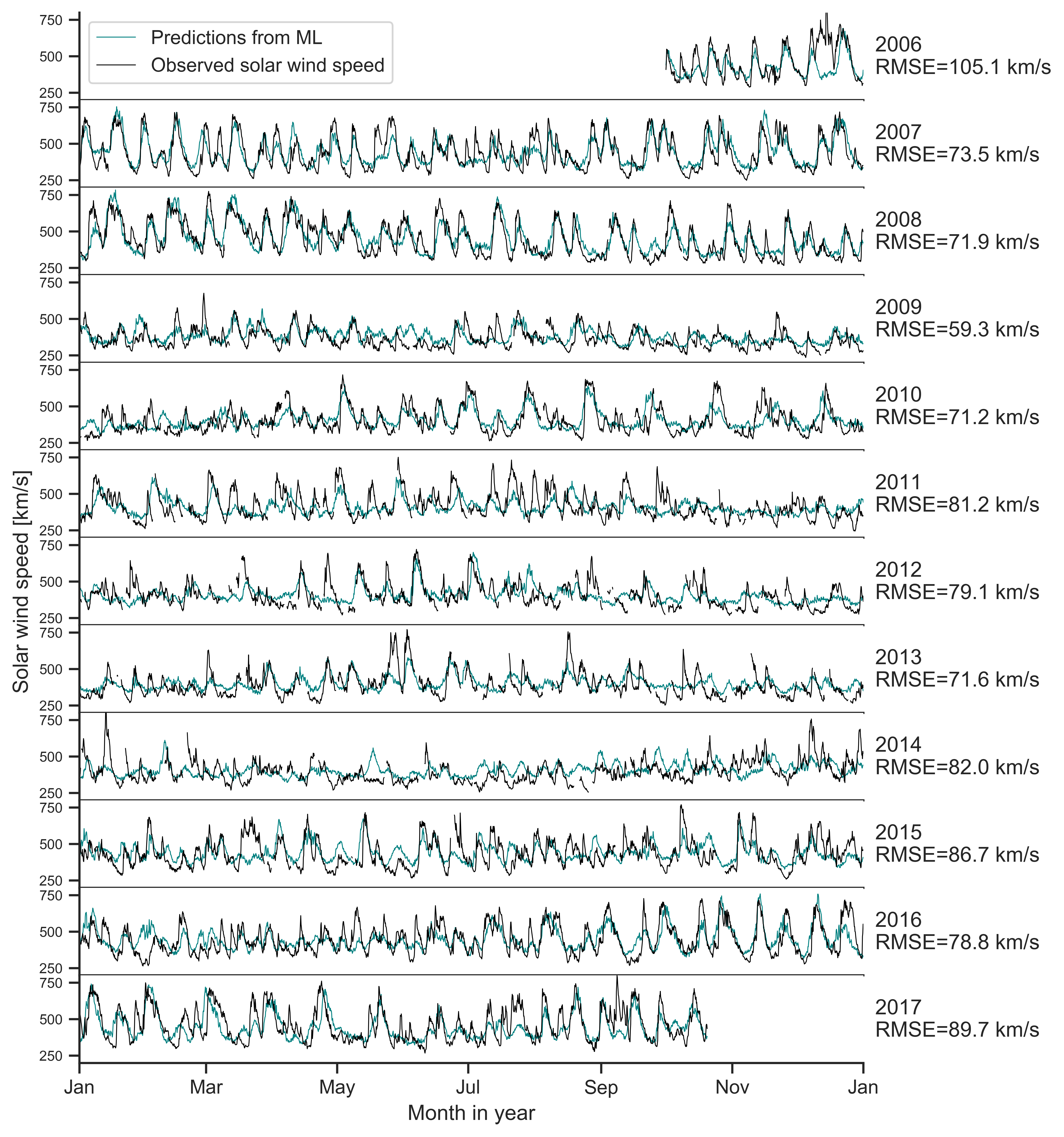}
\caption{From top to bottom: consecutive years of solar wind observations at Earth (black line) plotted alongside predictions from the machine learning model (teal). The RMSE for each year is added on the right. All y-axes are scaled to the range 200--800 km/s.}
\label{fig:pred_all_years}
\end{center}
\end{figure*}

\begin{figure*}
\begin{center}
\includegraphics[width=0.9\textwidth]{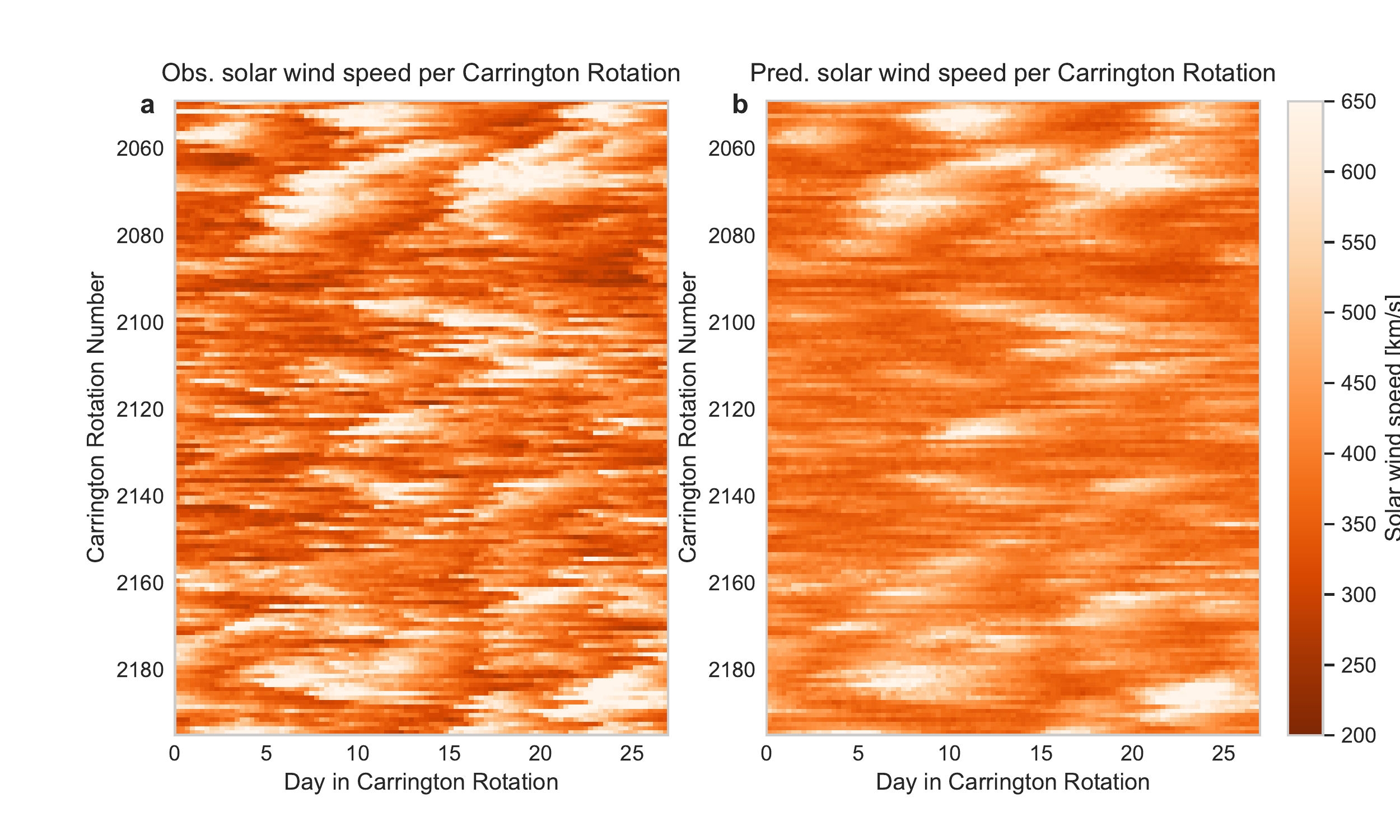}
\caption{The development of solar wind speed in observations and predictions over time. (a) and (b) depict the change in solar wind speed per Carrington rotation (27 days) for the entirety of the testing dataset, 11 years or 146 Carrington rotations. Time is increasing from top to bottom, and each line of the image is one Carrington rotation. (a) shows the solar wind speed in the observations at L1, while (b) shows the predicted solar wind speed from the machine learning model. As can be seen, recurrent patterns match well between observations and predictions, although short-term signals as seen mostly around solar maximum (middle of images) are not reproduced by the model.}
\label{fig:cr_speed_implot}
\end{center}
\end{figure*}

A validation of results for one solar cycle from the machine learning model with common metrics such as point-to-point measures shows that the GBR outperforms the WSA/HUX model combination and 27-day persistence in almost all measures. The results of the model are plotted alongside example input in Fig. \ref{fig:fp_d_time} for five Carrington rotations in Solar Cycle 24, which the model was tested on. Observed solar wind evolution is plotted in black in the lower panel (with solar wind speed determined by the WSA/HUX model in orange for comparison), and we see good agreement. The model applied to the entire test data set is plotted in Fig. \ref{fig:pred_all_years}. From a different perspective, figure \ref{fig:cr_speed_implot} shows the output of the model over the full temporal range of the test data as the development over all 146 Carrington rotations. The good visual agreement between the observations (left) and predictions (right) shows that the model approximates the ambient solar wind flows well. Shorter, transient solar wind flows that are seen primarily around solar maximum (middle of images) do not appear in the model predictions, but this is as expected as the model only predicts the evolving ambient solar wind.

Tables~\ref{tab:1}, \ref{tab:2} and \ref{tab:3} summarize the results obtained from model validation in terms of point-to-point measures, skill measures and event-based metrics, respectively. Multiple machine learning models with different input variables are considered, and the input variables are listed in the brackets. Besides the flux tube expansion factor $f_\text{p}$ and distance to the coronal hole boundary $d$, the solar wind speed from 27 days ago ($v_\text{pers}$) was also included.
We find that the best results in terms of established error functions are obtained with the GBR model based on $f_\text{p}$, $d$ and $v_\text{pers}$ with an RMSE of $78.3~\si{km.s^{-1}}$ and a PCC of $0.63$. (For reference, the RMSE and PCC values on the training / validation sets for this model were $96.1$ / $99.1$ km/s and $0.66$ / $0.64$, respectively. The higher RMSE is a consequence of SC 23 being far more active, and the same higher value is seen in the WSA model.) In comparison, the RMSE for the traditional WSA/HUX model combination is $98.9~\si{km.s^{-1}}$ and the PCC is $0.49$. The results for the model of 27-day persistence with an RMSE of $98.2~\si{km.s^{-1}}$ and a PCC of $0.52$ indicate that it provides a strong benchmark. The persistence model greatly benefits from having the same statistics as the observations as indicated by the similar AM and SD, and the repeating nature of the ambient solar wind flow. We also find that all the machine learning approaches investigated in this study improve the results in comparison to the climatological mean, as indicated by an SS of $0.02$, and an SS greater than $0.15$ for all the GBRs. 

\begin{table*}[t]
\caption{Classic point-to-point error measures for model comparison. Table of ambient solar wind prediction error metrics in terms of arithmetic mean (AM), standard deviation (SD), mean error (ME), mean absolute error (MAE), root mean square error (RMSE), the skill score (SS) relative to the climatological mean, and the Pearson correlation coefficient (PCC).}
\begin{center}
\begin{tabular}{lcccccccccc}
Model                   & {AM}      & SD        & ME        & MAE       & RMSE      & SS        & PCC\\ 
         & {[}\si{km.s^{-1}}{]} & {[}\si{km.s^{-1}}{]} & {[}\si{km.s^{-1}}{]} & {[}\si{km.s^{-1}}{]} & {[}\si{km.s^{-1}}{]} & \\\hline
WSA/HUX			        & 386.1     & 80.5      & 34.9      & 72.3      & 98.9      & 0.02      & 0.49      \\
GBR($d$)		        & 430.7     & 57.4      & -9.7      & 70.6      & 89.4      & 0.20      & 0.47      \\
GBR($f_\text{p}$)		& 429.1     & 49.6      & -8.1      & 72.6      & 91.3      & 0.16      & 0.42      \\
GBR($f_\text{p},d$)		& 427.7     & 61.6      & -6.7      & 66.8      & 85.0      & 0.28      & 0.54      \\
GBR($f_\text{p},d,v_{\text{pers}}$).  
                        & 422.3     & 71.6      & -1.3      & 59.9      & 78.3      & 0.39      & 0.63      \\
Persistence (27-days) 	& 420.8     & 99.8      & 0.1       & 71.9      & 98.2      & 0.03      & 0.52      \\
Observation  		    & 421.0     & 99.9      & -         & -         & -         & -         & - \\
\end{tabular}
\end{center}
\label{tab:1}
\end{table*}

In terms of the contingency table in Table \ref{tab:2}, we find that the application of the best machine learning approach improves all the skill measures, and particularly that the GBRs based on different feature combinations improve the number of hits and decrease the number of misses. While the TP count is quite high in GBR($d$) in comparison to the WSA/HUX model, the FP count is also high. GBR($f_\text{p}$), on the other hand, appears more conservative in predictions with lower TP and FP and with a larger number of FN. Combining the two in GBR($f_\text{p},d$) leads to improvements in all measures, and adding the $v_{\text{pers}}$ leads to yet more improvement. In comparison to the WSA/HUX, the machine learning techniques increase the number of false alarms and decrease the number of correct rejections. The TSS is greater than the value of $0.28$ (WSA/HUX model) for all the machine learning approaches. This positive increase in the TSS indicates that the overall performance of the machine learning techniques shows a positive trend towards correctly predicted enhanced solar wind conditions. While the traditional WSA/HUX model combination tends to under-predict the number of periods of enhanced solar wind speeds (BS=$0.57$), the machine learning approaches much better resemble the number of events in the solar wind observations. 


\begin{table*}[t]
\caption{Contingency table with skill measures of solar wind speed events. The event threshold is set at $v>450$ \si{km.s^{-1}}. The table shows the number of hits ({true positives}; TPs), false alarms ({false positives}; FPs), misses ({false negatives}, FNs), correct rejections ({true negatives}, TNs), followed by metrics derived from these values: the true positive rate (TPR) and false positive rate (FPR). The last three entries in each row show the threat score (TS), true skill statistics (TSS), and bias (BS).}
\begin{center}
\begin{tabular}{lccccccccc}
Model         & TP & FP & FN & TN  & TPR & FPR & TS & TSS & BS \\\hline
WSA/HUX	                & 3106    & 1711    & 5306    & 16410    & 0.37    & 0.09    & 0.31    & 0.28    & 0.57 \\
GBR($d$)	            & 4440    & 3722    & 3972    & 14399    & 0.53    & 0.21    & 0.37    & 0.32    & 0.97 \\
GBR($f_\text{p}$)	    & 3800    & 3118    & 4612    & 15003    & 0.45    & 0.17    & 0.33    & 0.28    & 0.82 \\
GBR($f_\text{p},d$)     & 4377    & 2969    & 4035    & 15152    & 0.52    & 0.16    & 0.38    & 0.36    & 0.87 \\
GBR($f_\text{p},d,v_\text{pers}$)       
                        & 4714    & 2368    & 3698    & 15753    & 0.56    & 0.13    & 0.44    & 0.43   & 0.84 \\
Persistence (27-days)	& 4873    & 3464    & 3472    & 14546    & 0.58    & 0.19    & 0.41    & 0.39    & 1.00 \\
\end{tabular}
\end{center}
\label{tab:2}
\end{table*}

\begin{figure*}
\begin{center}
\includegraphics[trim=5cm 5cm 5cm 5cm, width=0.3\textwidth]{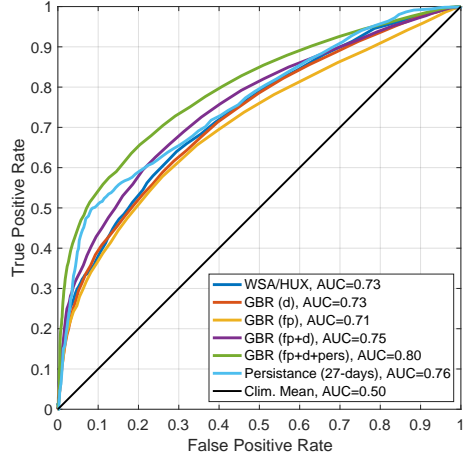}
\caption{Receiver operator characteristic curves showing the true positive rate (TPR) versus the false positive rate (FPR) for the investigated machine learning techniques and the reference models. The AUC (area under the curve) is included in the legend. The closer this value is to 1.0, the better the model performs.}
\label{fig:roc}
\end{center}
\end{figure*}

Figure~\ref{fig:roc} shows the computed ROC curves for all the solar wind modeling approaches investigated. We find that the GBR based on $f_\text{p}$, $d$ and $v_\text{pers}$ outperforms the WSA/HUX model combination for the full spectrum of selected event thresholds and outperforms persistence at almost all thresholds.

The last evaluation was an event-based analysis, shown in Table \ref{tab:3}. We find that the machine learning model slightly improves the results of the WSA/HUX model combination. We also find that the number of correctly predicted events increases for all the machine learning techniques and the number of misses decreases. However, the number of false alarms increases for the machine learning techniques. While the TS for the WSA/HUX model is $0.33$, the TS for the GBR based on $f_\text{p}$, $d$ and $v_\text{pers}$ is $0.40$. This indicates that most machine learning approaches provide a reasonable improvement in comparison to the traditional approach. However, the model of persistence again provides a challenging comparison in terms of an event-based validation analysis. 


\begin{table*}[t]
\caption{Statistics on event-based error metrics. An evaluation on the detections of high-speed enhancements was carried out in terms of event-based metrics including the number of observed (P) and forecast ($\text{P}_\text{F}$) events, the bias (BS), the number of hits (TPs), false alarms (FPs), and misses (FNs) together with the probability of detection (POD), false negative rate (FNR), positive predictive value (PPV), false alarm ratio (FAR), and threat score (TS).}
\begin{center}
\begin{tabular}{lccccccccccccccc}
Model    & P & P$_\text{F}$ & BS & TP & FP & FN & POD & FNR & PPV & FAR & TS \\ \hline
WSA/HUX	                & 451    & 223    & 0.49    & 167    & 56     & 284    & 0.37    & 0.63    & 0.75    & 0.25    & 0.33 \\
GBR($d$)		        & 451    & 305    & 0.68    & 201    & 104    & 250    & 0.45    & 0.55    & 0.66    & 0.34    & 0.36 \\
GBR($f_\text{p}$)	    & 451    & 279    & 0.62    & 179    & 100    & 272    & 0.40    & 0.60    & 0.64    & 0.36    & 0.33 \\
GBR($f_\text{p},d$)	    & 451    & 307    & 0.68    & 211    & 96     & 240    & 0.47    & 0.53    & 0.69    & 0.31    & 0.39 \\
GBR($f_\text{p},d,v_\text{pers}$)	   
                        & 451    & 292    & 0.65    & 213     & 79    & 235    & 0.48    & 0.52    & 0.73    & 0.27    & 0.40 \\
Persistence (27-days)	& 451    & 449    & 1.00    & 287    & 162    & 161    & 0.64    & 0.36    & 0.64    & 0.36    & 0.47 \\
\end{tabular}
\end{center}
\label{tab:3}
\end{table*}

\section{Discussion}\label{sec:discussion}


\subsection{Model Considerations}

There are some aspects to take into account when evaluating this work. First, the current modeling approach does not include any effects from the complex dynamic evolution of the ambient solar wind flow, and does therefore not consider any interactions between fast and slow ambient solar wind. Hence, it does not provide a picture of the solar wind conditions in the heliosphere and provides no self-consistent way to propagate the solutions from the Sun to the Earth. In order to account for the uncertainty in the time series, we make use of features based on time shifts at a cadence of $3.64$ hours ranging from 2 to 6 days. During the process of identifying the most important features, we find that those between 3-4 days are most critical, as would be expected from the typical transit time of ambient solar wind flows. 

Second, the model includes similar sources of inaccuracies as other models based on photospheric magnetic field measurements, such as the WSA model, and errors in the coronal magnetic model will extend to the predictions at L1. One error source is e.g. from the magnetic connectivity, when that the field line tracing reveals an incorrect source location of the solar wind. This can result from uncertainties in the magnetograms or the emergence of active regions on the far side of the Sun distorting the global field. Error sources such as these in the coronal domain explain solar wind speed enhancements missed by the model, such as those in the beginning of 2017 in Fig. \ref{fig:fp_d_time}. More generally, there is also the ongoing discussion over the location of the source surface height, which often provides better model accuracy when it varies across the cycle, as discussed specifically for solar cycles 23 and 24 in \citeA{Lee2011} and \citeA{Arden2014}.

Third, we note that the results in the present study have been deduced from an operational perspective. This means that magnetic features were computed once per day. Since most of the published literature uses the magnetic features computed only once per Carrington rotation, we also tested this setting for all the investigated machine learning models. We find that the model quality in terms of the RMSE of the GBR using this new setting decreases by about 5 percent. The knowledge of this variation is important to put our results into context with existing studies. We note that the final model, which can be accessed online on GitHub, can therefore also be directly applied to operational numerical frameworks for predicting the ambient solar wind.

One point worth emphasizing is that we have differentiated between training, validation and testing data sets in term of solar cycles. We used a period covering the second half of solar cycle 22 and most of solar cycle 23 for training and running validation to deduce the best parameters and features for the specific machine learning technique. We presented our validation analysis results on the test data set, which covered a period lasting one average solar cycle length (11 years), ranging from the end of solar cycle 23 and including most of solar cycle 24. None of the error estimates given refer to the output of the model on the data it was trained on, as doing so would not give an accurate estimate of the model's predictive skill when provided with new data. The results are therefore a reasonable estimate of the true-world performance of the regressor presented. In this context, we have also tried to exchange the roles of solar cycle 23 and 24, and found that the results for a model trained on solar cycle 24 and tested on solar cycle 23 are comparable, with no large variations in the error metrics. As an example, the RMSE for the testing on solar cycle 23 is $92.1$ km/s. Solar cycle 23 was considerably more active than cycle 24, and so a higher RMSE is to be expected. This highlights that the deduced regressor is not only fast but also robust. 

The machine learning approach presented here might improve the results of the established models such as WSA, but should not be considered a replacement for this modeling approach. For example, model combinations based on MHD include inherently important information on the location, boundaries and dynamics of open magnetic field lines along which high-speed solar wind streams accelerate into interplanetary space. Such information is essential as it ultimately improves our understanding of the underlying physical processes. 

The machine learning approach will be made available to the space weather community via the Ambient Solar Wind Validation Team embedded within the COSPAR/ISWAT initiative (\url{https://iswat-cospar.org/H1-01}). The computation time of the final machine learning product is very fast, requiring around 100 $\mu s$ (on a Macbook Pro 13'' A1708) to provide a set of 4-day predictions from data extracted from solar magnetic maps. This speed makes ensemble runs possible, with 1 million runs requiring 100 s - in an optimised setting on a dedicated server with multiprocessing it would only be a matter of seconds. One possible application could include uncertainties in the coronal field solutions in the solar wind speed predictions by running ensembles of all magnetic model features within $\pm 3^\circ$ of the sub-Earth track.
In the future, we shall work on several topics related to the improvement of the approach and on the extension thereof to other physical properties such as magnetic polarity, and magnetic sector boundary crossings. Furthermore, we plan on computing the global solar wind solutions near the Sun based on similar methodology. 

\subsection{Evaluating what didn't work}\label{sec:badfeatures}

One benefit of machine learning algorithms, particularly efficient methods such as gradient boosting regressors, is the ability to check and optimize for many different input variables rather than settling for a few. In this work, we presented models trained on three variables ($f_\text{p}$, $d$ and $v_\text{pers}$) but far more variables were tried before being excluded, and the discussion of which did and didn't work is interesting in its own right.

First, we included features related to the solar activity, with the F10.7 or MgII indices being widely applied examples of such. We expected the predictions to be improved when provided information on the activity level on the Sun. Surprisingly, models trained with either or both were found to perform no better on new data. This implies that sufficient information with regards to the state of the solar wind had already been provided via the variables used. In the cases of F10.7, the models were also more easily overtrained, implying that this index does not generalise well. This lends further confidence to our model's ability to correctly extract information on magnetic activity and open field cycle components from the features provided, without merely learning specific configurations.

Second, we investigated whether including details on the coronal holes as defined by the coronal models could lead to an improvement in the model accuracy. Details such as the area of coronal holes across all latitudes / low latitudes / high latitudes (taken across Earth-facing slices as with the other variables), average longitude of coronal holes, and total field in the coronal holes were included. The only variable that contributed to model accuracy was the area of coronal holes at low latitudes, while the others were negligible. A model trained on $f_\text{p}$ and $d$ with this coronal hole feature in addition saw an increase in accuracy of a few percent. This is in line with current literature and the results from with empirical prediction models based on EUV information, which perform best when dominant low-latitude coronal holes are observable, but they lack accuracy related to the temporal evolution of the solar wind \cite{reiss16}.  While the coronal hole feature did improve the accuracy of a model with $f_\text{p}$ and $d$, there was no increase when $v_\text{pers}$ was included, and to keep the analysis concise, we restricted ourselves to models trained on $f_\text{p}$ and $d$.

An investigation into the usage of ADAPT realizations is also interesting in this context. A model trained with all 12 ADAPT realizations was more easily overtrained, but a model trained with only one realization did not score as well in accuracy. Our analysis showed that taking three realizations proved a good balance, with various combinations resulting in models with similar levels of accuracy - the combination described in the results was most stable. The ADAPT ensemble represents the uncertainty of the flux location on the poles and far-side as a result of the random walk of flux driven by supergranulation flow patterns. If the ensemble is large enough to account for the major sources of uncertainty, then the “real solar wind” will be within the ensemble of solutions. It is a matter of ongoing research to try to find the best way to combine the many solutions, and what this study does is present a first approach - it is up to future work to optimise it further.

\subsection{Interpretability of machine learning models}

\add{Machine learning models, specifically deep learning models, often suffer from criticism of being difficult to interpret. The gradient boosting regression model is somewhat interpretable. One part that is easier to evaluate is the importance of each input feature determined by the proportion of branching points they make up, which depicts the predictive power of each variable.}

\add{During our analysis, we evaluated the roles of $f_\text{p}$ and $d$, which are connected to different theories on the origin of the slow solar wind. While the idea of $f_\text{p}$ is more related to waves and turbulence, $d$ implies that interchange reconnection at the coronal hole boundary determines the origin of the slow solar wind} \cite{Riley2015}\add{. In this context we studied the change in the feature importance of each variable over time. To do so, we trained models over shorter time periods (1-3 years) and evaluated how the feature importances change over the solar cycle. A pattern of duality between $f_\text{p}$ and $d$ emerges. We found that they alternated throughout the solar cycles, with $d$ providing better forecasts during the rising phase of the solar cycle, and $f_\text{p}$ during the declining phase. This feature importance analysis in the realm of machine learning offers an interesting perspective on this pending problem, but we cannot make a final statement on which of the variables is more important.}

\subsection{Comparison to similar studies}

We note that a similar approach already exists in the scientific literature, that by \citeA{Yang2018}. The study was carried out in the following way: 9 years of data were randomly split into training, validation and test data set, and another year of separated data was used for testing. This differs from our method, which has the test data separated out completely in time, and this was chosen in part because initial analyses on models trained and tested on data from the same solar cycle had lower errors but were found to not perform as well on new data. \citeA{Yang2018} have also used similar input data, including a 27-day persistence variable included as a feature. A direct comparison of the results is primarily difficult because of how the training and testing data sets have been defined in this study. Contrary to their use of a more complex neural network, we decided to use a GBR because of the fast computation times and simpler implementation paired with strong options for detailed model interpretation. In this way, it provides an efficient method that can be easily adapted in the future, such as being trained on new solar cycle data.

A recent study, \citeA{Chandorkar2020}, approaches the same problem with a more complex machine learning approach based on neural nets. In our study, uncertainty in the time it takes solar wind to travel from the Sun to the Earth has been accounted for by including all possible time windows from 2 to 5 days. \citeA{Chandorkar2020} instead built a \textit{dynamic time lag regression} algorithm around this timing problem, allowing the method to learn and account for time lag without an explicitly defined ground truth. Another difference is the time resolution: their model works with hourly forecasts, ours has a resolution of 3.64 hours. In terms of accuracy, their paper presents preliminary results for a defined time period, Carrington rotation \#2077, with MAE and RMSE values of $60.19$ km/s and $72.64$ km/s. The values for our model for the same time range come out at $45.19$ km/s and $56.75$ km/s. The comparison shows that our model performs better, but as we can only compare with one Carrington rotation this does not represent a final comparison of the two models.

\section{Summary}\label{sec:summary}

In this study we have presented an approach for predicting the ambient solar wind conditions in the vicinity of Earth. Specifically, we propose replacing the static WSA relation at the inner boundary of the heliospheric model domain by a machine learning technique, directly associating what is happening close to the Sun to what is happening near the Earth. We find that the features that are commonly employed in empirical techniques for specifying solar wind conditions near the Sun, namely the flux tube expansion factor $f_\text{p}$ and the distance to the coronal hole boundary $d$, are the most decisive features. Although in the first part we focus on improving the accuracy, our present investigation offers a new research tool for tackling the pending problem of ranking the relative importance of $f_\text{p}$ and $d$ and opens promising research avenues for the future. 

This is the first time that ADAPT realizations were combined with a machine learning method, offering a method for optimizing their usage. While it is possible to use all ADAPT realizations to predict the ambient solar wind flow, we find that a combination of a few realizations provides the best results.

The final model provides a forecast of the ambient solar wind with a lead time of 4.5 days in an ideal case, in which only data within 60 degrees west of the solar meridian is taken to eliminate possible inaccuracies resulting from lack of imaging and model projection at greater longitudes. This lead time could be extended to 9 days with a mission at the Sun-Earth Lagrange point 5 ($-60^\circ$ from the Earth) or 14 days if one entire half of the solar magnetic map from -180$^\circ$ to the meridian 0$^\circ$ is used.

With the presented results, we conclude that the machine learning technique, for which the code is provided open-source, enables a robust and fast approach to predict the solar wind speed near the Earth. The algorithm is easily transferable to other solar wind frameworks and is therefore an important contribution to the space weather community and can serve as a benchmark for future development of numerical ambient solar wind models. In a broader context, this study lays the foundation for future work on this subject, which will look into improving the modeling of solar wind conditions near the Sun \cite<see e.g.,>{Yang2019} and provide important input for MHD codes.

\section{Data Availability} \label{sec:data}

The materials used in this work are as follows:
\begin{itemize}
    \item ADAPT magnetic maps: \url{https://www.nso.edu/data/nisp-data/adapt-maps/}
    \item WSA coronal magnetic model maps: anyone interested in acquisition of this data should contact C.N.A. The data set can not be provided openly due to the competitive nature of the model. Alternatively, coronal magnetic maps from other models providing the same output could be used, and an example input data format is provided with the model code.
    \item OMNI data set for solar wind speed observations: \url{https://spdf.gsfc.nasa.gov/pub/data/omni/low_res_omni/}
    \item Jupyter Notebook used to train the model and produce the plots in this paper: \url{https://doi.org/10.5281/zenodo.4454989} or \url{https://github.com/helioforecast/Papers/tree/master/Bailey2021_AmbSoWiML}
    \item OSEA was used for statistical analysis: \url{https://doi.org/10.5281/zenodo.3753104}
\end{itemize}

\acknowledgments
R.L.B., M.A.R., C.M., U.V.A., T.A., A.J.W. and J.H. thank the Austrian Science Fund (FWF): P31659-N27, J4160-N27, P31521-N27, and P31265-N27. M.A.R. would like to thank NASA's CCMC for travel support. Part of this work was carried out during a research stay at NASA Goddard in December 2019. Details on where to find the data used in this study as well as the development code are given in Section \ref{sec:data}. We would also like to thank the editor and anonymous reviewers for their help in improving this work.


\bibliography{bib}

%
%
%
%
%

\end{document}